\documentclass[conference]{IEEEtran}
\usepackage{graphicx}
\usepackage{amsmath,amssymb}
\usepackage{url}
\usepackage{dcolumn}
\usepackage{adjustbox}
\usepackage{multirow}
\usepackage{flushend}

\newenvironment{definition}[1][Problem Definition]{\begin{trivlist}
\item[\hskip \labelsep {\bfseries #1}]}{\end{trivlist}}

\begin{document}

\title{Network Backboning with Noisy Data}

\author{\IEEEauthorblockN{Michele Coscia}
\IEEEauthorblockA{Center for International Development\\
Harvard University\\
Cambridge, MA 02138\\
Email: michele\_coscia@hks.harvard.edu}
\and
\IEEEauthorblockN{Frank M. H. Neffke}
\IEEEauthorblockA{Center for International Development\\
Harvard University\\
Cambridge, MA 02138\\
Email: frank\_neffke@hks.harvard.edu}
}

\maketitle

\begin{abstract}
Networks are powerful instruments to study complex phenomena, but they become hard to analyze in data that contain noise. Network backbones provide a tool to extract the latent structure from noisy networks by pruning non-salient edges. We describe a new approach to extract such backbones. We assume that edge weights are drawn from a binomial distribution, and estimate the error-variance in edge weights using a Bayesian framework. Our approach uses a more realistic null model for the edge weight creation process than prior work. In particular, it simultaneously considers the propensity of nodes to send and receive connections, whereas previous approaches only considered nodes as emitters of edges. We test our model with real world networks of different types (flows, stocks, co-occurrences, directed, undirected) and show that our Noise-Corrected approach returns backbones that outperform other approaches on a number of criteria. Our approach is scalable, able to deal with networks with millions of edges.
\end{abstract}

\IEEEpeerreviewmaketitle

\section{Introduction}\label{sec:introduction}
In this paper, we present a new algorithm to extract the significant edge backbone from a noisy complex network.

Scientists studying complex phenomena have discovered networks as a powerful tool for their analysis \cite{albert2002statistical}. Examples are the discovery of functional modules in networks \cite{coscia2013structure}, the prediction of missing connections \cite{liben2007link} and the modeling of information cascades \cite{cha2009measurement}. Such tools have been applied to a vast and diverse set of applications, ranging from cultural analysis \cite{park2015topology} to viral marketing \cite{chen2009efficient} and the improvement of the efficiency of road networks \cite{zheng2016keyword}.

However, these applications face a number of challenges. One among the most prevalent ones concerns data quality. In recent years, there has been an exponential increase in production and recording of human data. This increase has enabled the success of large scale network analysis in the first place. However, the combination of data from disparate sources with different measurement techniques and biases means that connections retrieved from these data are noisy, yielding networks in which many nodes have a large degree such that the underlying structure becomes difficult to parse, because all entities connect to each other.

This is the scenario that network backboning algorithms address. A network backboning algorithm is an algorithm that takes as input a dense and noisy network and returns a reduced version of it. The reduction is performed with the aim of highlighting the underlying structure of the network, removing spurious connections. This network backbone can now be used to provide a clearer image of the phenomenon the network relates to, that can be studied by standard network analysis tools, such as community discovery, link prediction, or other network algorithms.

\begin{figure}
\centering
\includegraphics[width=.72\columnwidth]{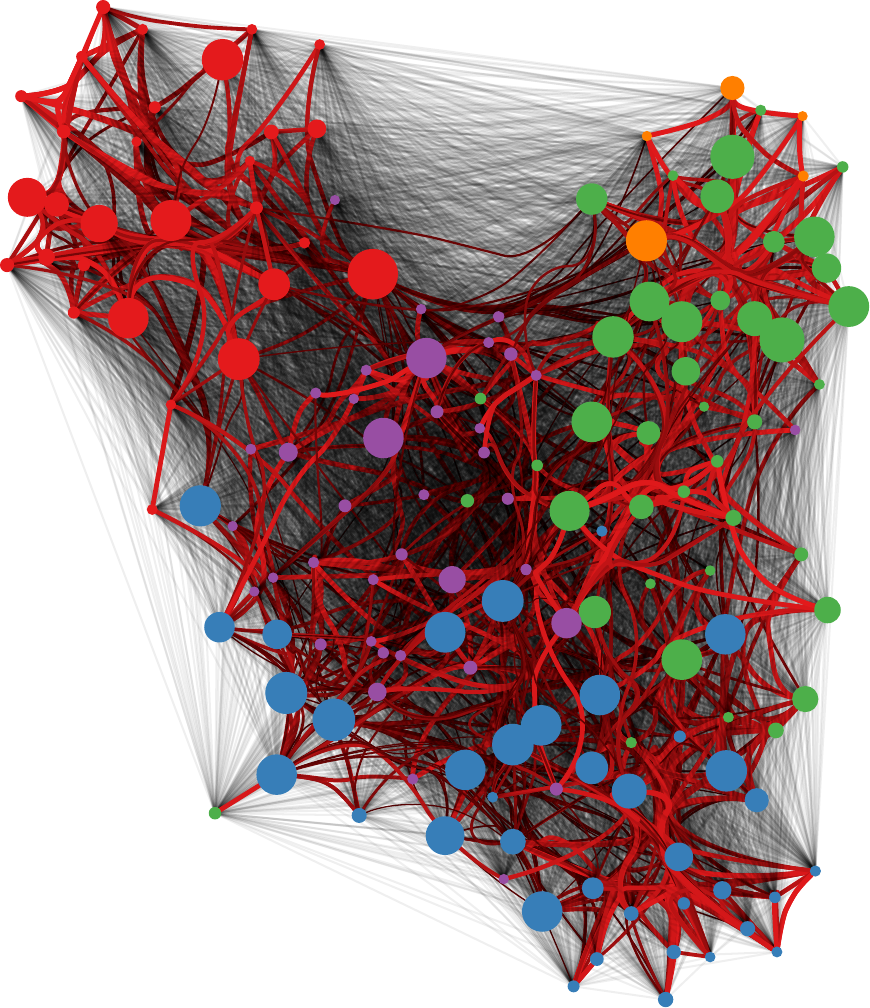}
\caption{An example of network backbone.}
\label{fig:backbone-example}
\end{figure}

Consider Figure \ref{fig:backbone-example} as an example. This is a network with only 151 nodes, where virtually every possible connection is expressed in the data. However, the vast majority of these connections are weak and do no represent a significant interaction between the nodes. In other words, these edges reflect noise. The noisy connections are represented as thin black lines. A network backboning algorithm aims at evaluating the significance of each edge weight for the two nodes sharing the connection. In the figure, the backbone is represented by the wide red edges that were determined to be statistically significant. Once we prune noisy edges from the network, the underlying structure emerges more clearly, and it can be used as input of network analysis algorithms. In this case we used a community discovery algorithm that attempts to retrieve the ground truth classes of the nodes, represented by the node color. Such task cannot be performed in the original network, because the density of connections leads the community discovery algorithm to classify all nodes into the same giant community.

As an example, we will consider the task of predicting inter-occupational flows of jobs-switchers. A way to predict these flows is to build a network that connects occupations that require common skills and tasks. However, certain skills are so generic that they show up in most occupations, leading to spurious connections. That is, the likelihood of having such skills in common is large, even for random pairs of occupations. A good backboning algorithm will drop these random connections. The resulting backbone will then be a better predictive tool for our task than the unfiltered network.

Current approaches to network backboning can be classified into two categories: generalist and specialized. In the specialized class, the algorithm focuses on solving a specific problem and it optimizes the backbone to fit the particular problem definition. This paper provides a generalist backboning approach, where the algorithm is agnostic about the specific application. However, it is particularly suited for applications in which edge strengths have a count-data structure.

Generalist network backboning algorithms can use one of two approaches. In the first, the algorithm is defined structurally: the algorithm trusts that there is no noise in the edge weights and it prunes the edges according to a criterion that lets the latent structure of the network emerge. Prominent exponents of this class are the High Salience Skeleton (HSS) \cite{grady2012robust} and the Doubly-Stochastic Transformation (DST) \cite{slater2009two}. This paper belongs to the second category, where we want to assess the statistical significance of an edge weight to decide whether to prune the edge or not. The state of the art methodology in this class is the Disparity Filter \cite{serrano2009extracting}. In the Disparity Filter (DF), edge weights are tested against a null model of the node emitting them: if an edge weight is significant compared to the total outgoing weights of the node, the edge is kept in the backbone.

In our approach, which we call \textit{Noise-Corrected} (NC) backbone, we assume edge weights are drawn from a binomial distribution. Then, we estimate the probability of observing a weight $N_{ij}$ connecting nodes $i$ and $j$ using a Bayesian framework. This framework enables us to generate posterior variances for all edges. This posterior variance allows us to create a confidence interval for each edge weight. In practice, we drop an edge if its weight is less than $\delta$ standard deviations stronger than the expectation, where $\delta$ is the only parameter of the NC algorithm. However, the confidence intervals the algorithm produces can also be used more generally, for instance to determine whether two edges differ significantly from one another in strength.

Our approach improves over DF in several respects. First, our null model does not only consider the propensity of the origin node to send connections, but also the propensity of the destination node to receive connections. In the DF, links connecting peripheral nodes to hubs are kept, because periphery-hub connections always seem strong from the peripheral node's perspective, even though the strong attraction of the hub makes it likely that such edges form randomly. However, the much less-anticipated peripheral-peripheral connections between nodes are more likely to be dropped in the DF backbone. The NC backbone preserves the latter connections at the expense of the former. Second, we will argue that we use a more realistic null model for edge weights than DF. Finally, the NC backbone outperforms the DF backbone in real world applications.

We provide evidence in favor of the last claim in the experimental section. We believe that a good backbone satisfies three properties. First, a good backbone must have a reasonable network topology. What is ``reasonable'' is hard to define but, arguably, it is desirable to prevent that nodes get isolated by dropping the entirety of their connections. NC satisfies the criterion in all networks by retaining, for a given number of edges, a high number of nodes under most circumstances. Second, a good backbone has to highlight the underlying properties of the original data, and therewith make modeling efforts easier. We show that NC backbones improve the predictive power of simple models in our network datasets, more so than any alternative. Finally, a good backbone should be stable, as the underlying phenomenon does not change drastically between $t$ and $t + 1$. The NC backbones have a stability equivalent to the one of DF backbones. These tests have been run on networks of countries, connected by: business travel, trade, similarity in export baskets and so on. We chose this application because country networks belong to different types (flows, stocks, co-occurrences) and have different topological properties (directed, undirected). Moreover, it is easier to evaluate the quality of a backbone since we have reasonable priors about which countries should be connected with which others. Finally, country networks are widely used in a variety of applications, e.g., economic development \cite{atlas} and aid \cite{coscia2013structure}. We start our analysis by showing on synthetic networks that the NC backbone recovers the original network in presence of large amounts of noise better than any alternative.

The NC backbone implementation is able to scale to millions of edges in less than two minutes. This shows the readiness of our method to be deployed to real world problems. The implementation of the NC backbone is freely available as a Python module\footnote{\url{http://www.michelecoscia.com/?page_id=287}}. To ensure result reproducibility, we also release some of the country networks used in this paper. Pursuant to contractual obligations we are unable to provide a release of the full dataset.

\section{Related Work}
Network backboning is a problem related to Principal Component Analysis (PCA) \cite{jolliffe2002principal}. PCA is used to reduce the dimensionality of a matrix: to decompose it in a smaller matrix that still represents the original data as well as possible. A network backbone aims at doing the same for network structures. The current network backboning approaches were recently reviewed in \cite{hamann2016structure}. The classic ways to do network backboning are: establishing a naive threshold to prune edges with low weights, calculate the maximum spanning tree, or applying k-core decomposition, where nodes with degree lower than $k$ are recursively removed from the network \cite{seidman1983network}.

We can divide network backboning methods into two classes: generalist and specialized. In the generalist class we have methods that reduce the dimensionality of the edge set to let the underlying structure of the network emerge \cite{guimera2009missing}. This is usually useful to simplify ``hairballs'' where the real structure (be it community structure rather than core-periphery) is hidden in noise. Under this category we find the Disparity Filter \cite{serrano2009extracting}, the Doubly-Stochastic transformation \cite{slater2009two}, and the High Salience Skeleton \cite{grady2012robust}. This is the literature we are addressing here, so these methods will be considered more in depth in Section \ref{sec:back-art}.

In the second class, we have methods that were developed with a very specific problem definition in mind. Here, the analyst is not interested in the overall general structure of the network hidden behind the noise, but in a specific application of an algorithm. For instance, one method sparsifies graphs for the task of inferring the influential nodes in a social network \cite{mathioudakis2011sparsification}, sometimes known as ``graph simplification'' \cite{bonchi2013activity}. In other cases, the backbone is discovered from the activities of nodes, rather than from the edge weights \cite{chawla2016discovering}. Related methods focus on graph compression, which also reduces the set of nodes for very large graphs \cite{toivonen2011compression}. Graph sampling \cite{leskovec2006sampling} can be used for local-first algorithms and also carried out on data streams \cite{ahmed2014network}. Finally, there has been some work on defining backbones in the emerging field of multilayer network analysis \cite{foti2011nonparametric}. Multilayer networks are networks in which there are multiple edge types \cite{berlingerio2011foundations, kivela2014multilayer}, and backbones have to also take the inter-layer couplings into account.

To find the backbone of a network is very often the first step of the entire network analysis framework. As such, the application scenarios of network backboning are many, and diverse. We can find network backboning methods used in online systems \cite{zhang2013extracting}, cultural analysis \cite{park2015topology}, microbial co-occurrence networks \cite{connor2016using}, query-dependent graph summarization \cite{shi2015vegas}, and to boost clustering \cite{satuluri2011local}.

\section{Background}

\subsection{Problem Definition}
The fundamental data structure in this paper is a weighted graph $\mathcal{G} = (\mathcal{V}, \mathcal{E}, N)$, where $\mathcal{V}$ is the set of vertices; $N \subseteq \mathbb{R}^{+}$ is the set of non-negative real edge weights; and $\mathcal{E}$ is a set of triples $(i,j,n)$ with $i,j \in \mathcal{V}$ and $n \in N$. Each triple represents an edge. Let $N_{ij}$  be the weight of the edge connecting nodes $i$ and $j$. Furthermore, let subscript ``$.$'' indicate a summation over the omitted dimension, i.e.: $N_{i.} = \sum_{j} N_{ij}$ is the total outgoing weights of $i$; $N_{.j} = \sum_{j} N_{ij}$ is the total incoming weights of $j$; and $N_{..} = \sum_{i,j} N_{ij}$ the total weights in the network. Edges can be directed -- meaning $(i,j,n) \neq (j,i,n)$ --, or undirected -- meaning $(i,j,n) = (j,i,n)$.

In this paper, we aim at solving the problem of extracting the backbone from a dense complex network. A backbone is informally defined as a subset of nodes and edges of the original network that contains the largest possible subset of nodes and the smallest possible subset of salient connections. Formally:

\begin{definition}
Given a weighted graph $\mathcal{G} = (\mathcal{V}, \mathcal{E}, N)$, find a graph $\mathcal{G}^* = (\mathcal{V}^*, \mathcal{E}^*, N^*)$, such that $\mathcal{E}^* \subset \mathcal{E}$, and: 
\begin{itemize}
\item $\left| \mathcal{I}_{\mathcal{G}^*} \right| \approx \left| \mathcal{I}_{\mathcal{G}} \right|$, where $I_\mathcal{G}$ is the set of isolates in graph $\mathcal{G}$ (Coverage);
\item $f\left(\mathcal{E}^*,\mathcal{V}^* \right) >f\left( \mathcal{E},\mathcal{V} \right) $, where $f\left(. \right)$ is the fit of a prediction task (Quality);
\item $D\left(\mathcal{E}^*_1, \mathcal{E}^*_2,\right) < D\left(\mathcal{E}_1, \mathcal{E}_2,\right)$, where $\mathcal{G}_1$ and $\mathcal{G}_2$ are two independent measurements of $\mathcal{G}$ and $D\left(. \right)$ a distance metric (Stability).
\end{itemize}
\end{definition}

\subsection{State of the Art}\label{sec:back-art}
In this section we present the network backboning algorithms with which we compare our proposed methodology more in depth. We focus exclusively on methods for generic backbone extractions, which are used more often in network science. The main purpose of our method, as we describe in Section \ref{sec:method}, is an accurate estimation of the amount of noise in the edge weights. Once noise is estimated, one can filter out the edges whose weight is not sufficiently different from the random benchmark, thus obtaining a noise-reduced backbone.

Facing the problem of filtering connections in dense networks, the simplest approach is to remove all edges whose weight is lower than an arbitrary threshold $\delta$. We call this the ``Naive'' approach. The Naive approach has several downsides. In most real-world weighted networks, weights are broadly distributed, locally correlated on edges incident to the same node, and nontrivially coupled with topology \cite{serrano2009extracting, barrat2004architecture}. This is true also for the networks we use in Section \ref{sec:exp}. Broad distributions imply the lack of a characteristic scale, rendering the choice of $\delta$ meaningless; local correlations imply the possibility of discarding valuable information by filtering out entire parts of the network. For these reasons, the network science community started developing alternative methodologies.

One such methodology is the calculation of the so-called ``Maximum Spanning Tree'' (MST). A MST is the spanning tree of a connected, undirected graph whose total weighting for its edges is maximum. A spanning tree is a tree connecting all vertices of the graph. If $\mathcal{T}$ is the MST of $\mathcal{G}$, $\nexists \mathcal{T}' s.t. \sum N^{*}_{\mathcal{T}'} > \sum N^{*}_{\mathcal{T}}$. Maximum spanning trees are usually extracted with the Kruskal algorithm \cite{kruskal1956shortest}. Given its definition, it is easy to see that the MST satisfies the first condition of our backbone extraction problem definition: it preserves all nodes and there is no way to select the same number of edges with a higher sum of weights. However, maximum spanning trees have several shortcomings. First, if there are repeated edge weights in the graph, there could be multiple MSTs. Second, by definition, a MST is a tree, which might omit fundamental properties of real world networks such as transitivity and communities \cite{boccaletti2006complex}.

An alternative to Naive and MST backboning was developed with a two-step algorithm \cite{slater2009two}. In the first step the adjacency matrix of the network is transformed into a doubly-stochastic matrix by alternatingly normalizing the row and column values with their respective sums. Then edges are sorted in descending normalized weight. Edges are added to the backbone from strongest to weakest, until the backbone contains all original nodes in $\mathcal{V}$ in a single connected component. We call this the ``Doubly-Stochastic'' (DS) method. The DS method has three downsides. First, it requires the adjacency matrix to be square, thus excluding bipartite networks. Second, it is not always possible to transform any arbitrary square matrix into a doubly-stochastic one \cite{sinkhorn1964relationship}. Third, it provides no theoretical foundation for dealing with noisy weight estimations.

Two of the most used algorithms recently developed in network science are the Disparity Filter \cite{serrano2009extracting} and the High Salience Skeleton \cite{grady2012robust} (HSS). The HSS is defined structurally. For each node, \cite{grady2012robust} calculates the shortest path tree (SPT) connecting the node to all other nodes in the network. Then, the HSS is the linear superposition of all SPTs, i.e. $HSS = \sum \limits_{v \in \mathcal{V}} SPT(v)$. Also the HSS does not explicitly estimate the noise in the edge weight estimation, and thus noisy edges could degenerate the final HSS.

The Disparity Filter (DF) determines whether an edge carries a significant share of the total edge weights that enter or depart from a given node. It does so node by node. The significance of an edge is hence determined with respect to a single node, as opposed to a node-pair. Edges are first expressed as shares of a node's total incoming edge weights. Next, these shares are compared to a null-model, yielding p-values for how far a share exceeds its null-model's prediction. The null-model assumes that the distribution of node's $k$ edge shares are generated by a random process where $k-1$ points are randomly drawn from a uniform distribution. These points divide the unit interval into $k$ pieces. The length of each piece now represents one of the edge weights. Note that this is equivalent to assuming that edge weights are drawn from an exponential distribution. For each edge, the filter determines how likely it is that the $k-1$ other edges would leave sufficient edge-weight-share for the edge to assume the weight it has or more. This likelihood acts as a p-value for how significantly an edge weight deviates from the null-model prediction. These p-values are then used to prune insignificant edges. In practice, an edge is tested twice to verify whether its weight is significant for either of the connected nodes as emitters or receivers. However, the null-model ignores the interaction between source and target node. 

\section{Methods}\label{sec:method}
We refer to our method as Noise-Corrected (NC) backbone. The method is based on \cite{neffke2016coordinated}, which uses the null-model described here to deal with attenuation biases that arise from measurement errors in the context of regression analysis. When comparing to the state of the art in network backboning, the NC backbone most closely relates to the Disparity Filter. As in the DF, we compare edge weights to a null model, but we formulate this null model at the level of node pairs, not individual nodes. The NC is constructed in three steps. First, we transform edge weights such that they are expressed in deviation from their null-model prediction. Second, we calculate a standard deviation for these transformed edge weights. Third, we use these standard deviations to construct p-values that are then used to prune edges.

Let us think of edge weights as the sum of unitary interactions that occur with constant, edge-specific, probability $P_{ij}$. Furthermore, take the total interactions emitted and received by a node (i.e., a node's summed incoming or outgoing edge weights) as given.  Using ``\^{}'' to denote observed quantities, the expected number of interactions in node pair $(i,j)$ can be written as:

$$E[N_{ij}]=\hat{N}_{i.} \frac{\hat{N}_{.j}}{\hat{N}_{..}}$$

That is, we assume that each interaction that starts from node $i$ finds destination node $j$ with a probability that equals the share of total interactions in the network received by $j$. We compare the observed to expected edge weights to calculate an edge weight's lift \cite{witten2005data}:

$$ L_{ij}=\frac{\hat{N}_{ij}}{E[N_{ij}]}$$

The lift tells us how unexpectedly high an edge weight is given the weights of $i$ and $j$. When lift equals one it means that the edge has the expected weight. Values from one to infinity indicate a stronger than expected connection, from one to zero a weaker than expected connection. As one can see, the lift is an asymmetric skewed measure, where a value of 0.1 is equally far from one on the left as 10 is from the right. We therefore transform $L_{ij}$, using the following monotonic mapping:

\begin{equation}
\tilde{L}_{ij}=\frac{L_{ij}-1}{L_{ij}+1}=\frac{\kappa\hat{N}_{ij}-1}{\kappa\hat{N}_{ij}+1}
\label{eq:lift}
\end{equation}

where $\kappa=\frac{1}{E[{N}_{ij}]}$. The transformation in Eq. \ref{eq:lift} ensures that the lift becomes a symmetric measure centered on zero. In our example, .1 becomes equal to $-0.\overline{81}$ and 10 becomes equal to $0.\overline{81}$.

We now need to compute the variance of $\tilde{L}_{ij}$, which is given by:

$$V\left[\tilde{L}_{ij}\right]=V\left[\frac{\kappa\hat{N}_{ij}-1}{\kappa\hat{N}_{ij}+1}\right].$$

Applying the delta method, we get:

\[
V\left[\tilde{L}_{ij}\right]=V[\hat{N}_{ij}]\left(\frac{2\left(\kappa+\hat{N}_{ij}\frac{d\kappa}{d\hat{N}_{ij}}\right)}{\left(\kappa\hat{N}_{ij}+1\right)^{2}}\right)^{2}
\]

with

\[
\frac{d\kappa}{d\hat{N}_{ij}}=\frac{1}{\hat{N}_{i.}\hat{N}_{.j}} -\hat{N}_{..}\frac{\hat{N}_{i.}+\hat{N}_{.j}}{\left(\hat{N}_{i.}\hat{N}_{.j}\right)^{2}}.
\]

We now have to estimate $V[\hat{N}_{ij}]$. Given that we have interpreted edge weights as the sum of independent unitary interactions, $N_{ij}$ follows a Binomial distribution with variance:

\begin{equation}
V\left[N_{ij}\right]=N_{..}P_{ij}\left(1-P_{ij}\right) \label{eq:varNij}
\end{equation}

$P_{ij}$ is unknown, but can be estimated as the observed frequency with which interactions occur:
$$\hat{P}_{ij}=\frac{\hat{N}_{ij}}{\hat{N_{..}}}.$$

A problem arises when $\hat{N}_{ij} = 0$. That is, when edge weights are zero for certain node pairs. Given that many real-world networks are sparse, this situation is quite common. For these node pairs, $V\left[N_{ij}\right] = 0$, which would suggest that measurement error is absent in these edges. However, in reality, there is simply too little information to estimate $\hat{P}_{ij}$ with sufficient precision. This affects not only cases where edge weights are zero, but also where information is sparse, i.e., when focusing on the interactions among nodes of low degree. To improve on this, we estimate $\hat{P}_{ij}$ in a Bayesian framework. That is:

\noindent
\resizebox{\columnwidth}{!}{
$ 
Pr\left[N_{ij}=n_{ij}\left|N_{..}=n_{..},P_{ij}=p_{ij}\right.\right]=\left(\begin{array}{c}
n_{..}\\
n_{ij}
\end{array}\right)p_{ij}^{n_{ij}}\left(1-p_{ij}\right)^{n_{..}-n_{ij}}
$
}

Using Bayes' law, we get:

\noindent
$
Pr\left[P_{ij}=p_{ij}\left|N_{..}=n_{..},N_{ij}=n_{ij}\right.\right]=
$

\noindent
\resizebox{\columnwidth}{!}{
$
\frac{Pr\left[N_{ij}=n_{ij}\left|N_{..}=n_{..},P_{ij}=p_{ij}\right.\right]Pr\left[P_{ij}=p_{ij}\left|N_{..}=n_{..}\right.\right]}{\int_{0}^{1}Pr\left[N_{ij}=n_{ij}\left|N_{..}=n_{..},P_{ij}=q_{ij}\right.\right]Pr\left[P_{ij}=q_{ij}\left|N_{..}=n_{..}\right.\right]dq_{ij}}
$
}
\begin{equation}
\label{eq:bayes}
\end{equation}

Choosing a $BETA\left[\alpha,\beta\right]$ distribution, the conjugate prior of the Binomial distribution, as a prior for $P_{ij}$, the posterior distribution is also a $BETA$ distribution. In particular, the posterior distribution of $P_{ij}$ becomes:

\begin{equation}
P_{ij}\sim BETA\left[n_{ij}+\alpha,n_{..}-n_{ij}+\beta\right]. \label{eq:posterior}
\end{equation}

We still have to choose values for $\alpha$ and $\beta$ that would give plausible prior expectations for the mean and variance of $P_{ij}$. To do so, assume that the total weight of $i$ and $j$ is given.  In other words, think of edge weights as arising from a process in which, each time node $i$ increases its total weight by one, it draws a node $j$ at random from the pool of possible nodes. That is, edge weight generation follows a hypergeometric distribution. This gives the following prior means and variances for $P_{ij}$:

$$E\left[P_{ij}\right]=E\left[\frac{N_{ij}}{N_{..}}\right]=\frac{1}{N_{..}}E\left[N_{ij}\right]:=\frac{1}{N_{..}}\frac{N_{i.}N_{.j}}{N_{..}}$$

$$V\left[P_{ij}\right]=\frac{1}{N_{..}^{2}}V\left[N_{ij}\right]:=\frac{1}{N_{..}^{2}}\frac{N_{i.}N_{.j}\left(N_{..}-N_{i.}\right)\left(N_{..}-N_{.j}\right)}{N_{..}^{2}\left(N_{..}-1\right)}.$$

The $:=$ equality indicates where we make our assumptions. From the $BETA\left[\alpha,\beta\right]$ distribution, we get:

\begin{equation}
E\left[p_{ij}\right]=\mu=\frac{\alpha}{\alpha+\beta}
\label{eq:E}
\end{equation}

\begin{equation}
V\left[p_{ij}\right]=\sigma^{2}=\frac{\alpha\beta}{\left(\alpha+\beta\right)^{2}\left(\alpha+\beta+1\right)}
\label{eq:V}
\end{equation}

Solving for $\alpha$ and $\beta$, we get:

\begin{equation}
\alpha=\frac{\mu^{2}}{\sigma^{2}}\left(1-\mu\right)-\mu
\label{eq:alpha}
\end{equation}

\begin{equation}
\beta=\mu\left(\frac{\left(1-\mu\right)^{2}}{\sigma^{2}}+1\right)-1
\label{eq:beta}
\end{equation}

Eqs. \ref{eq:posterior}, \ref{eq:E}, \ref{eq:V}, \ref{eq:alpha} and \ref{eq:beta} now define a posterior expectation for $P_{ij}$  for each node pair. We can use this posterior expectation of $P_{ij}$ instead of $\hat{P}_{ij}$ to recalculate variances of edge weights in Eq. \ref{eq:varNij}. Because the posterior expectation of $P_{ij}$ is always strictly larger than zero, variance estimates do not degenerate.

\begin{figure}
\centering
\includegraphics[width=.46\columnwidth]{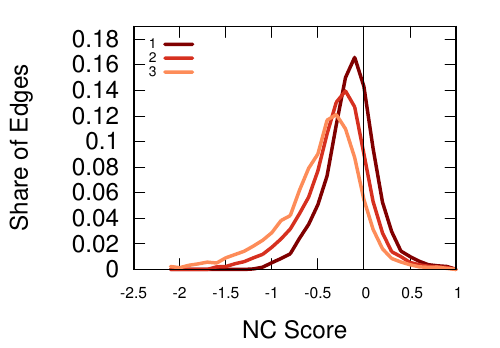}
\includegraphics[width=.46\columnwidth]{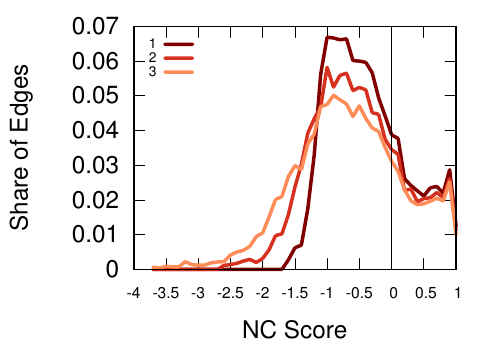}
\caption{Two examples of threshold setting for the Country Space (left) and Business (right) networks, which will be presented in Section \ref{sec:exp-data}.}
\label{fig:threshold}
\end{figure}

At this point we have an estimation of the expected variance of the edge weight $\tilde{L}_{ij}$: $V\left[c_{ij}\right]$. An edge is kept if and only if its observed weight is higher than $\delta \sqrt{V\left[\tilde{L}_{ij}\right]}$. i.e. it surpasses the expectation by $\delta$ standard deviations. $\delta$ is the only parameter of the NC backbone. It has to be set accordingly to the tolerance to noise that the particular application can have. To visualize the effect of $\delta$, consider Figure \ref{fig:threshold}. Here we show the distribution of $\tilde{L}_{ij} - \delta \sqrt{V\left[\tilde{L}_{ij}\right]}$ for different $\delta$ values (one to three). Higher $\delta$s shift the distribution to the left. The vertical black bar at zero shows the boundary between accepted and rejected edges. The acceptance area lies to the right. Since this is roughly equivalent to a one-tailed test of statistical significance, common values of $\delta$ are 1.28, 1.64, and 2.32, which approximate p-values of 0.1, 0.05, and 0.01\footnote{An alternative approach, implemented in the python package, is to skip the transformation step described above. p-values follow directly from the null-model's Binomial distribution, using $N_{..}$ as the number of draws and $\frac{\hat{N}_{i.}\hat{N}_{.j}}{\hat{N}_{..}^{2}}$ as probability of success. However, in this way we are not able to estimate the standard deviation of the estimate. This makes impossible, for instance, to determine if two edges are significantly different from each other.}.

\begin{figure}
\centering
\includegraphics[width=.6\columnwidth]{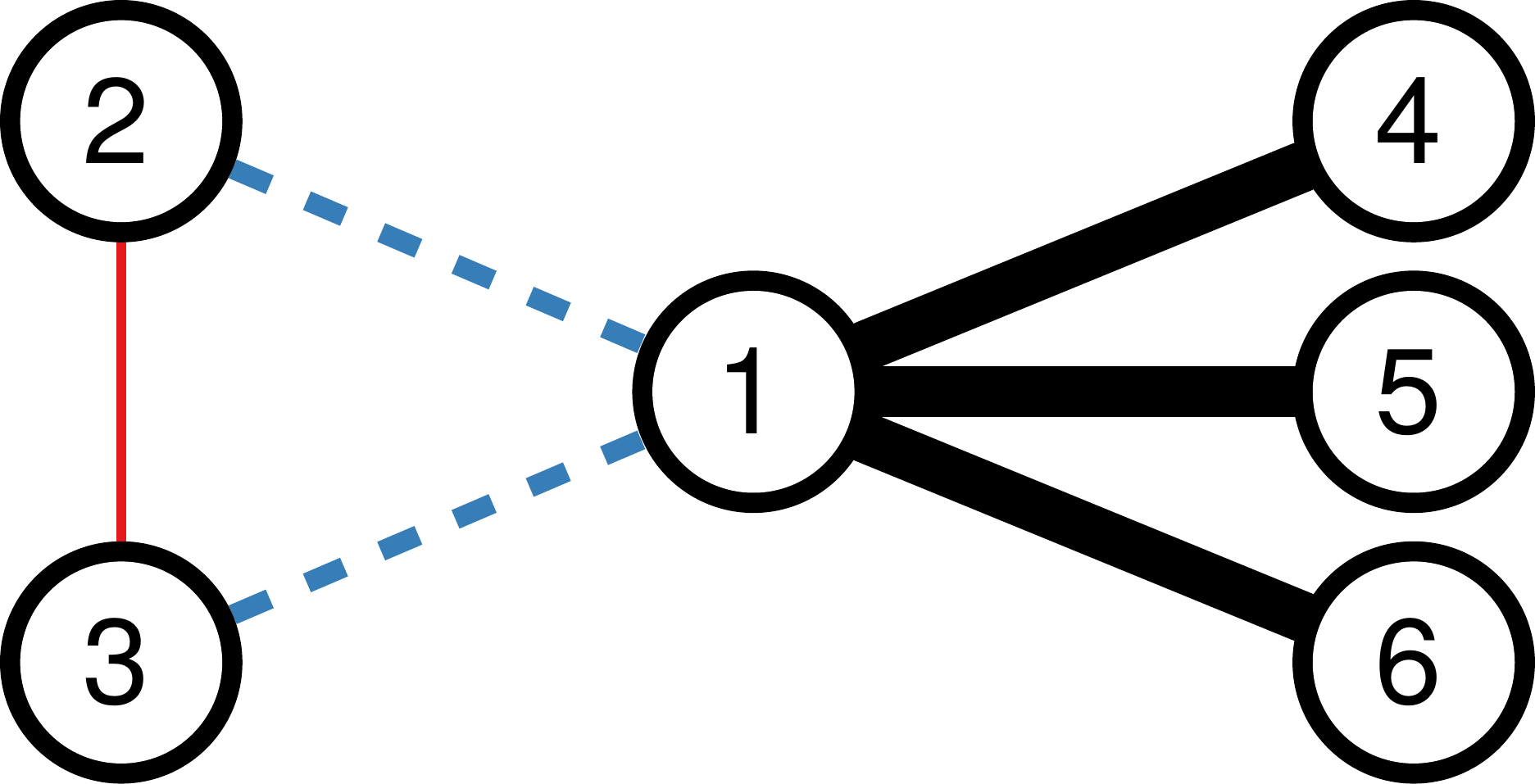}
\caption{A toy example of the difference between the Noise-Corrected and the Disparity Filter backbones.}
\label{fig:toy-neffke-vesp}
\end{figure}

Now that we have formally defined our method, we can directly compare it to the Disparity Filter on a toy example. Consider Figure \ref{fig:toy-neffke-vesp}. Here we have a hub connected to five nodes. Two of them are also connected to each other. The edges are undirected and their width is proportional to their weight. The black edges are selected by both the DF and our NC backbones. The DF backbone furthermore selects the blue dashed edges connecting the hub (node 1) to the peripheral nodes. From the perspective of the hub these edges should be dropped, given that they are expected to arise even under random edge formation, but when considering the perspective of the other nodes as emitters, the strength of the connections to the hub makes them highly unanticipated. On the other hand, the NC backbone finds the connection between nodes 2 and 3 more important than connections to node 1. That is because connecting to node 1 is not extraordinary, given its propensity to connect to everything, and to dispense high weighted edges. However, even though the connections 1-2 and 1-3 are stronger than 2-3, the latter is more unanticipated, because nodes 2 and 3 tend to have low edge weights in general. Hence, the fact that these weak nodes connect to each other strongly suggests that this edge represents a deviation from randomness.

We publicly release our implementation as the {\tt backboning} Python module. The module contains also the implementations of the Disparity Filter \cite{serrano2009extracting}, the High Salience Skeleton \cite{grady2012robust}, the Doubly Stochastic transformation \cite{slater2009two}, as well as Maximum Spanning Tree and naive thresholding. These are the methods to which we compare the NC backbone in the following experiment section. The module is available at the link provided in Section \ref{sec:introduction}.

\section{Experiments}\label{sec:exp}
In this section we perform a series of experiments to show the effectiveness of the Noise-Corrected backbones. We start by building synthetic networks with added noise, showing how the NC backbone is able to recover the original edge set in presence of high levels of noise (Section \ref{sec:exp-synthetic}). Then, we focus on real-world networks. The data comes from a series of country-country networks presented in Section \ref{sec:exp-data}. We start by showing that our edge weight variance estimation correlates with the actual variance of the edge weight as observed in the real world, validating our assumed null model (Section \ref{sec:exp-validation}). Then, we compare the NC method with the state of the art over three criteria of success. As stated in the problem definition, a good backbone is a backbone that: (a) produces the largest possible node coverage of the network (\textbf{Topology} -- Section \ref{sec:exp-topology}); (b) improves the usefulness of the network data for prediction tasks (\textbf{Quality} -- Section \ref{sec:exp-quality}); and (c) reduces topology fluctuations in time (\textbf{Stability} -- Section \ref{sec:exp-stability}). We validate the scalability of our approach in Section \ref{sec:exp-scalability}.

\subsection{Synthetic Networks}\label{sec:exp-synthetic}
In this section we test the performance of each method in recovering the backbone of synthetic networks. In this scenario, we have perfect information about which edge is part of the actual network and which edge reflects noise. We generate several Barabasi-Albert random networks with average degree 3 and 200 nodes. We set $\eta$ as our noise parameter. Each actual edge in the Barabasi-Albert network carries the following weight:

$$ N_{ij} = (k_i + k_j) \times \mathcal{U}(\eta, 1), $$

where $k_i$ is the degree of node $i$, and $\mathcal{U}(\eta, 1)$ is a number extracted from a uniform distribution with minimum $\eta$ and maximum 1. In practice, we use a fraction of at least $\eta$ of the sum of the degrees of the connected nodes. In this way, we ensure broad edge weight distributions locally correlated with the network topology. Then, the complement of the adjacency matrix is filled with noisy edge weights, which are defined with the same formula, only changing the uniform element with $\mathcal{U}(0, \eta)$. In practice, a noisy edge can have at most a fraction $\eta$ of the degrees of $i$ and $j$.

For all methods we set the parameters (if any) so that the backbone will return the same number of edges as the underlying actual network. Our quality target is the Jaccard coefficient between the set of edges of the original non-noisy network and the backbone. It is equal to one if the two edge sets are identical, and to zero if they do not share a single common edge.

\begin{figure}
\centering
\includegraphics[width=.72\columnwidth]{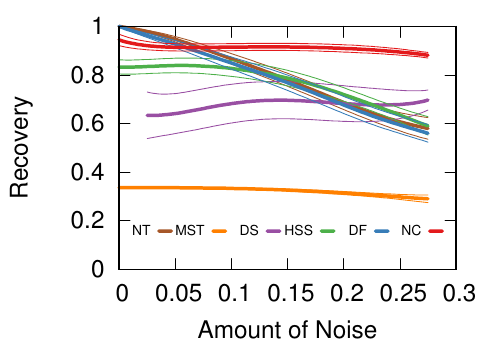}
\caption{Recovery of the true backbone of synthetic Barabasi-Albert networks.}
\label{fig:synthetic}
\end{figure}

Figure \ref{fig:synthetic} reports the results. For very low amount of noise, the two best performing solutions are the Disparity Filter and the naive thresholding. However, our Noise-Corrected backbone is more resilient to increasing noise with the best overall performance, while also performing very well in low-noise environments. As noise increases, there is no significant difference between DF and naive thresholds.

\subsection{Data}\label{sec:exp-data}
Our test set includes six networks. In all networks, nodes are countries and connections are relationships among them, calculated in six different ways. The six networks, in alphabetical and discussion order, are as follows.

\begin{itemize}
\item \textbf{Business}: two countries are connected through the number of corporate credit cards issued in one country (origin) that are utilized for expenditures in another (destination). This is a directed flow network, observed in the years 2012, 2013 and 2014. Anonymized and aggregated data from Mastercard’s Center for Inclusive Growth.\footnote{\url{http://mastercardcenter.org/}}.
\item \textbf{Country Space}: two countries are connected with the number of products they both export in significant quantities. This is an undirected co-occurrences network, observed in the years 2011, 2012 and 2013. Trade data comes from \cite{cepii_trade}. To determine whether exports are significant we use the same criterion of \cite{atlas}, based on the concept of Revealed Comparative Advantage.
\item \textbf{Flight}: two countries are connected through the existing passenger capacity in flights from airports in one country (origin) to another (destination). This is a directed flow network, observed in the years 2010 and 2014. Proprietary data from OAG\footnote{\url{http://www.oag.com/}}.
\item \textbf{Migration}: two countries are connected through the number of total migrants from one country (origin) currently living in another (destination). This is a directed stock network, observed in the years 1990, 2000, 2010 and 2013. Data from the UN \cite{un_migration}.
\item \textbf{Ownership}: two countries are connected through the number of total establishments in a country (destination) reporting to a global headquarter in a different country (origin). This is a directed stock network, observed in the years 2008, 2011 and 2014. Proprietary data from Dun \& Bradstreet\footnote{\url{http://www.dnb.com/}}.
\item \textbf{Trade}: two countries are connected through the total dollar value of all products exported by one country (origin) and imported by another (destination). This is a directed flow network, observed in the years 2011, 2012 and 2013. Trade data have been cleaned with the same procedure outlined in \cite{atlas}.
\end{itemize}    

Additional edge attribute tables used in prediction tasks record: the distance between two countries as the weighted average distance between all pairs of major cities  in these countries \cite{cepii_geo}; population data for all countries from the World Development Indicators of the World Bank\footnote{\url{http://data.worldbank.org/indicator/SP.POP.TOTL}}.

\begin{figure}
\centering
\includegraphics[width=.72\columnwidth]{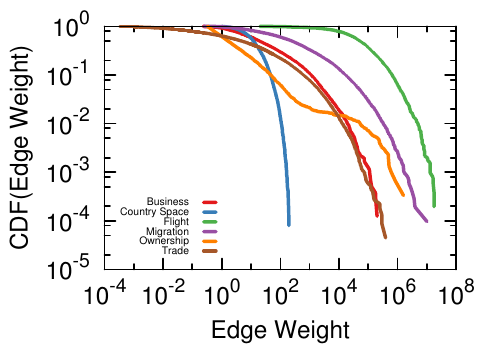}
\caption{Cumulative edge weight distributions for our networks.}
\label{fig:weights}
\end{figure}

\begin{figure*}
\centering
\includegraphics[width=.3\textwidth]{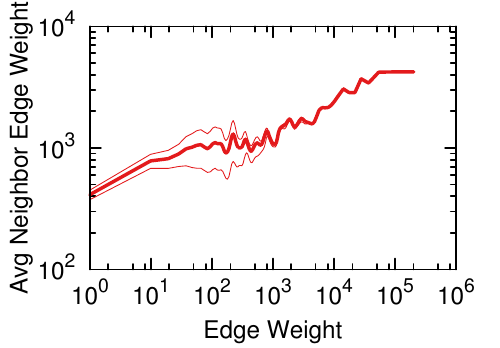}
\includegraphics[width=.3\textwidth]{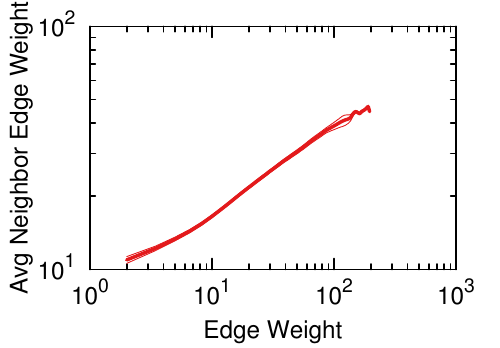}
\includegraphics[width=.3\textwidth]{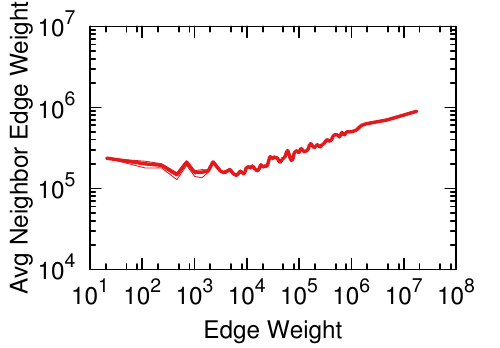}\\
\includegraphics[width=.3\textwidth]{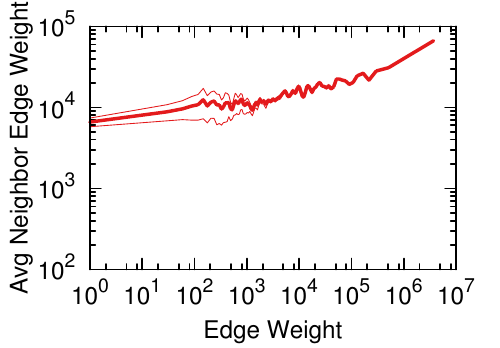}
\includegraphics[width=.3\textwidth]{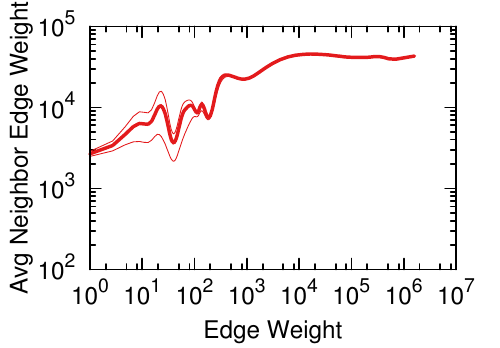}
\includegraphics[width=.3\textwidth]{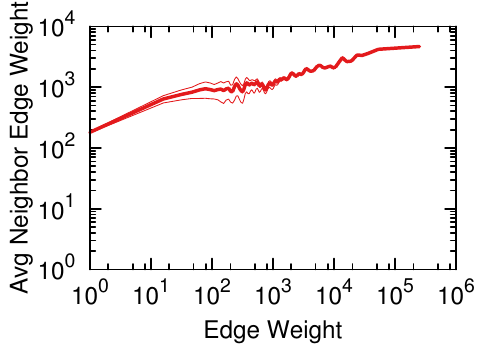}
\caption{Edge weight vs average neighboring edge weight. From left to right: (top) Business, Country Space, Flight, (bottom) Migration, Ownership and Trade.}
\label{fig:neighweights}
\end{figure*}

One of the main reasons for using a sophisticated backboning method instead of a naive threshold is the broad and locally correlated distribution of edge weights. This has been observed in many weighted networks \cite{barrat2004architecture}, and it is also the case in the country-country networks used here. We report the cumulative edge weight distributions of all six networks in Figure \ref{fig:weights}. Figure \ref{fig:weights} shows that all the studied networks have broad degree distributions (although neither of them is a power-law), with the possible exception of the Country Space network. For instance, in the Ownership network the median non-zero edge weight is 1.5, while the top 1\% of non-zero edges have weights larger than 50k. The Trade network edge weights span ten orders of magnitude.

The edge weights are also locally correlated. We calculate the log-log Pearson correlation between the weight of an edge and the average weight of the edges connected to either of its nodes. Figure \ref{fig:neighweights} plots the average and variance of neighbor edge weights against an edge's own weight. The correlation is weakest in the Flight network, but at a value of .42 still highly statistically significant ($p < 10^{-15}$). The strongest correlation -- in the Country Space network -- equals to .75.

\subsection{Validation}\label{sec:exp-validation}

\begin{table}
\centering
\begin{tabular}{l|r}
Network & NC Corr\\
\hline
Business & .590\\
Country Space & .627\\
Flight & .613\\
Migration & .064\\
Ownership & .872\\
Trade & .162\\
\end{tabular}
\caption{The correlation coefficients between predicted and observed variance for the NC and DF backbones.}\label{tab:variance-correlation}
\end{table}

Before looking at the quality of the actual backbone, we validate our methodology against real world data. The Noise-Corrected backbone aims at correctly estimating $V\left[\tilde{L}_{ij}\right]$, i.e. the variance of the transformed edge weights. Since we observe all the country networks in several points in time, we can compare our expectation of $V\left[\tilde{L}_{ij}\right]$ with the actual (observed) variance. 

Table \ref{tab:variance-correlation} reports the calculated correlation coefficients. All correlation coefficients are significant with $p < 10^{-9}$. 

\subsection{Topology}\label{sec:exp-topology}
\begin{figure*}
\centering
\includegraphics[width=\textwidth]{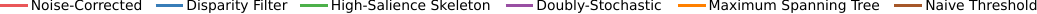}
\includegraphics[width=.3\textwidth]{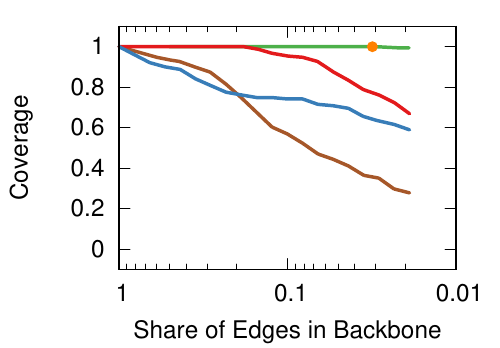}
\includegraphics[width=.3\textwidth]{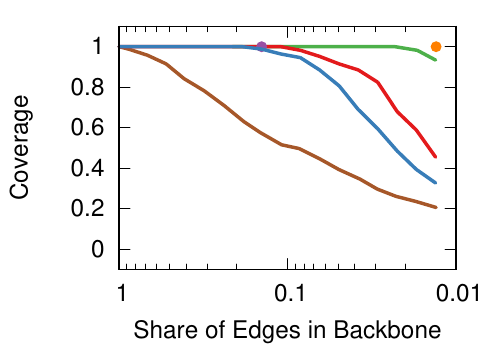}
\includegraphics[width=.3\textwidth]{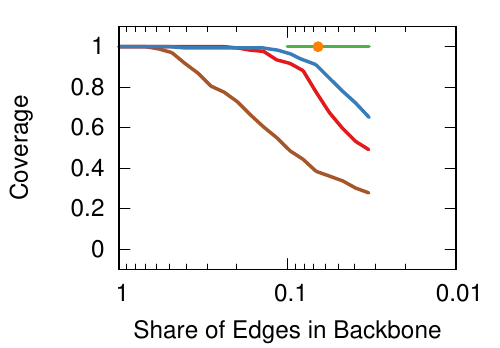}\\
\includegraphics[width=.3\textwidth]{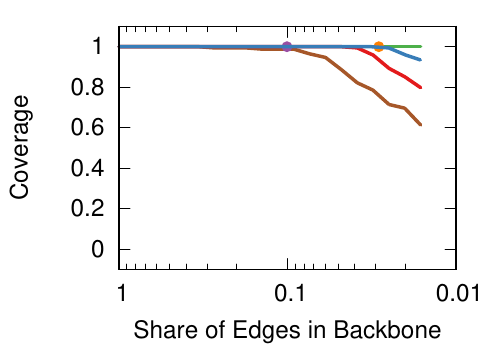}
\includegraphics[width=.3\textwidth]{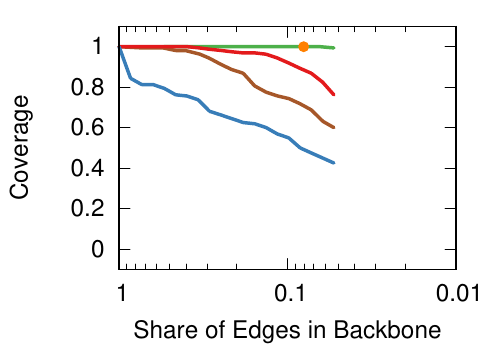}
\includegraphics[width=.3\textwidth]{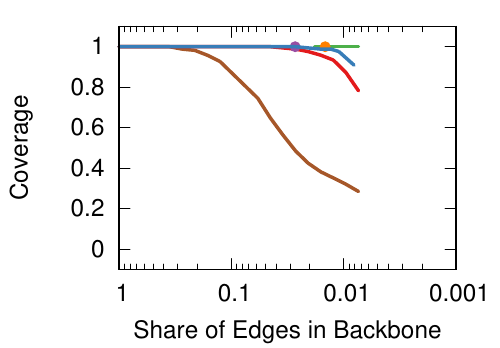}
\caption{Coverage per backbone for varying threshold values. From left to right: (top) Business, Country Space, Flight; (bottom) Migration, Ownership, Trade.}
\label{fig:nodes}
\end{figure*}

As the first quality criterion, we focus on the topology of the backbone. In general, backbones ideally isolate as few nodes as possible: each node dropped by the backbone is a node for which we will not have any result from the network analysis. Thus, we define the Coverage as the ratio between non-isolated nodes in the backbone over non-isolated  nodes in the original network, or:

$$ Coverage = \dfrac{|\mathcal{V}|-|I_{G}^{*}|}{|\mathcal{V}|-|I_{G}|}.$$ 

Figure \ref{fig:nodes} reports the number of nodes preserved in each backbone as a function of the share of edges that was preserved. Note that the Doubly Stochastic method is present only for the networks Country Space, Migration and Trade, as for the other networks the stochastic transformation was not possible. Note that DS and MST do not require any parameter, so they appear in the plot as a point rather than as a line.

Many data points overlap because in many instances all methods were able to preserve the entirety of the node set. However, it is easy to detect the cases in which a particular method was not able to achieve perfect coverage. MST, DS and HSS achieve perfect coverage by definition (the latter fails only for very strict parameter choices). There is no clear winner between NC and DF, as in some networks one achieves better coverage than the other, while the converse is true for others. However, the DF is the only method underperforming the naive method in one case (the Ownership network), which is a critical failure.

\subsection{Quality}\label{sec:exp-quality}
In this section we argue that a good backbone lets the underlying properties of the data emerge from the noisy data. To prove this point we create a series of models for OLS regressions. These models all have the same structure:

$$\log(N_{ij} + 1) = \beta X_{ij} + \epsilon_{ij}.$$

\begin{table*}
\centering
\begin{tabular}{l|rrrrrr}
Method & Business & Country Space & Flight & Migration & Ownership & Trade\\
\hline
Doubly Stochastic & n/a & 2.0975 & n/a & 1.5153 & n/a & 0.9287\\
Naive Threshold & 0.7766 & 0.6834 & 0.5196 & 1.1616 & 1.2384 & 0.3935\\
Disparity Filter & 0.9315 & 1.4082 & 0.8569 & 2.0715 & 0.5374 & 0.9024\\
High Salience Skeleton & 1.1341 & 1.6549 & 0.9447 & 1.2597 & 0.9744 & 0.8662\\
Maximum Spanning Tree & 1.1183 & 1.9180 & 0.7981 & 1.0036 & 0.9288 & 0.9532\\
Noise-Corrected & \textbf{1.1767} & \textbf{2.2437} & \textbf{1.4676} & \textbf{2.1493} & \textbf{1.4165} & \textbf{1.1037}\\
\end{tabular}
\caption{The improvement in predictive power when using backbones in our six networks.}
\label{tab:quality}
\end{table*}

Here, $N_{ij}$ is the edge weights of the network, $X_{ij}$ is a collection of variables that are supposedly good predictors of $N_{ij}$, and $\epsilon$ is the error term. In practice, we propose a model to explain the connection strength between countries. Then, for each backbone methodology we run two regressions. In the first regression $M_{full}$, we use the complete set of observations. In the second model $M_{bb}$, we restrict the observations edges that are contained in the network backbone. Quality is then defined as:

$$ Quality = \dfrac{R^2_{M_{bb}}}{R^2_{M_{full}}},$$

which is the ratio of the quality prediction (the $R^2$ of the OLS model) obtained using the backbone to restrict observations over the baseline quality that uses all edges. A value of 1 means that the two regressions have equivalent predictive power, while 
a value higher than 1 means we improve over the full network.

To allow for a fair comparison of different backbone methodologies, we fix the number of edges we include in the backbone. We usually choose the number of edges obtained with low threshold values for the High Salience Skeleton, because it is the most strict backbone methodology in our collection, always returning the fewest number of edges. Note that this does not apply to the MST and DS backbones, since they do not have parameters and thus the number of edges cannot be tuned.

\begin{figure*}
\centering
\includegraphics[width=\textwidth]{legend.pdf}
\includegraphics[width=.3\textwidth]{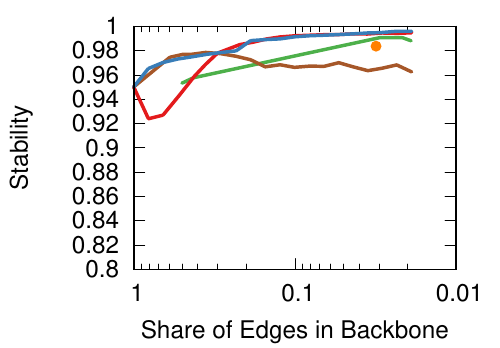}
\includegraphics[width=.3\textwidth]{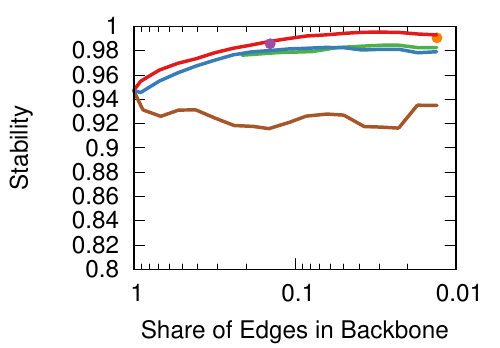}
\includegraphics[width=.3\textwidth]{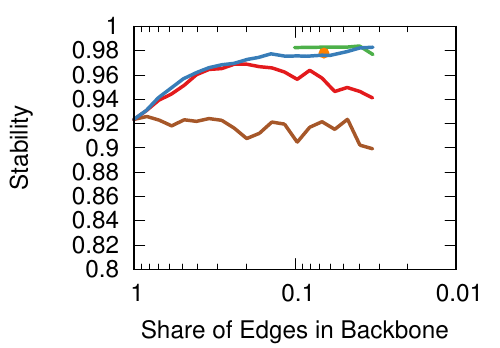}\\
\includegraphics[width=.3\textwidth]{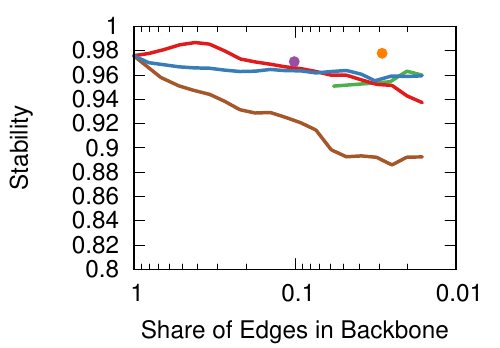}
\includegraphics[width=.3\textwidth]{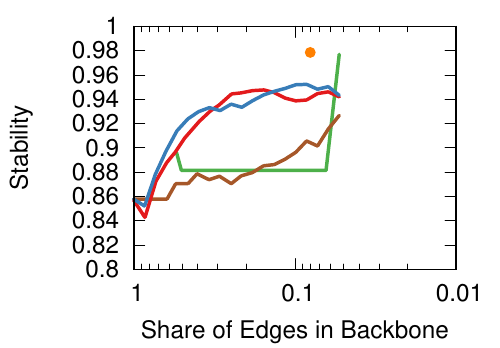}
\includegraphics[width=.3\textwidth]{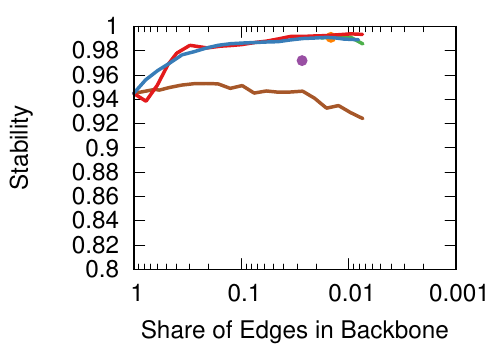}
\caption{Stability per backbone for varying threshold values. From left to right: (top) Business, Country Space, Flight; (bottom) Migration, Ownership, Trade.}
\label{fig:stability}
\end{figure*}

One predictor used in each network is geographical distance: nearby countries are assumed to be more connected. For almost all networks (except Country Space and Ownership) we also use population-size as a control: since we are recording flows and stocks, large countries are expected to have larger connections. Finally, we add the following network-specific predictors: \textbf{Business}: trade between the countries. Trading partners are expected to maintain a high intensity of business travels. \textbf{Country Space}: economic complexity (ECI \cite{atlas}) of the two countries. Two countries with similar technology levels are expected to export similarly advanced products. \textbf{Flight}: no additional variable. Country size and distance suffice for a typical gravity model: we expect many travelers between nearby large countries. \textbf{Migration}: cultural variables like common language and common history \cite{cepii_geo}. More migrants are expected if the destination country has the same language and customs as the origin country. \textbf{Ownership}: foreign direct greenfield investments\footnote{\url{https://www.fdimarkets.com/}} (FDIs). Each greenfield establishment in a foreign country has been created with FDI, thus total dollar investment between the countries should predict the number of establishments. \textbf{Trade}: business travels between countries. Two countries that visit each other for business are expected to trade frequently with each other.

Table \ref{tab:quality} reports the results. We highlight the best performing methodology in boldface. Note that in some cases the doubly stochastic transformation was not possible, thus we label these instances as ``n/a''. In all cases, the NC backbone performs better than any other backbone. What is more, the NC backbone is also the only method that always returns a quality value higher than one, meaning that the backbone outperforms the original, unfiltered, network in all cases.

\subsection{Stability}\label{sec:exp-stability}
Finally, we are interested in the stability of a backbone. Because most of our networks should be relatively stable, we consider wild year-on-year fluctuations in an edge's weight as a sign that the edge weight is imprecisely measured. Our backbone should contain fewer noisy edges and therefore be more stable than the orginal network. The stability of a backbone method can be calculated as:

$$ Stability = corr(N^{t}_{ij}, N^{t + 1}_{ij}),$$

where $corr$ is the Spearman correlation coefficient between the two vectors. In principle, any distance metric is appropriate for this task, but we prefer the nonparamatric nature of the Spearman coorelation, which mimics our task of ranking edges according to their significance. We calculate the correlation using only the edges present in the backbones. This means that a perfectly stable backbone will have a stability of 1, while a value of 0 implies that there is no relation between the backbones of time $t$ and $t + 1$. We calculate stability across different thresholds.

Figure \ref{fig:stability} depicts the results. There is no clear winner in this quality criterion. All backbones are very stable, with stability always exceding .84. 

\subsection{Scalability}\label{sec:exp-scalability}
\begin{figure}
\centering
\includegraphics[width=.72\columnwidth]{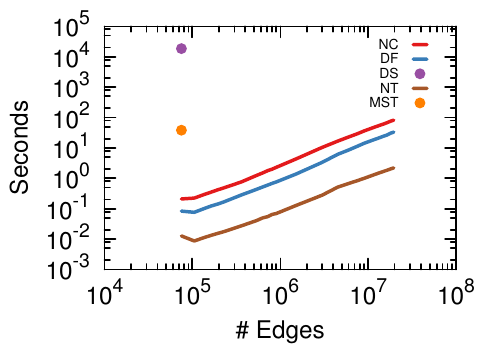}
\caption{Running time scalability.}
\label{fig:scalability}
\end{figure}

We implemented the NC backbone using the {\tt pandas}\footnote{\url{http://pandas.pydata.org/}} Python library. In this section we provide a quantification of the running time of this naive implementation, showing that it is ready for real world tasks. We generate a set of Erd\"{o}s-R\'{e}nyi graphs, with uniform random weights. We set the average degree of a node to three and we generate networks from 25 thousand to 6.5 million nodes. For each network size, we repeat the backbone computation ten times and report the average.

Figure \ref{fig:scalability} depicts the time complexity scaling in terms of number of edges. Running time increases only slightly superlinearly with the number of edges. Empirically, we estimate the time complexity of our implementation to be $\sim \mathcal{O}(|\mathcal{E}|^{1.14})$. The average running time of the algorithm for a network containing 20 million edges was 82 seconds on a Xeon(R) CPU E5-2630 at 2.30GHz.

For completeness, we compare with equivalent implementations of the other methods. The NC backbone scales with a rate indistinguishable from the linear naive thresholding and DF, which differ only by a constant multiplicative factor. Other methods such as HSS and DS are much less time efficient and we could not run them on networks larger than a few thousand edges.

\section{Case Study}
In this section we provide a more in-depth discussion of an application of the network backboning techniques. The aim is to show how an analysis task can be improved by filtering out noisy connections from a network. The topic we study is the estimation of skill relatedness \cite{neffke2013skill}. In \cite{neffke2013skill}, the authors connected cells in an industry space with the aim of capturing similarities in industry skill requirements, assuming that such similarities would be reflected in inter-industry labor flows.

Here we explore the validity of this assumption. We investigate the relationship between a skill-co-occurrence network and inter-occupational labor flows. Next, we restrict the connections, using a backboning technique and show how this improves the quality of predictions. There are a few differences between this case study and \cite{neffke2013skill} that should be noted. First, we use occupations instead of industries as nodes. Second, in \cite{neffke2013skill}, the authors use a confidential dataset from Sweden that cannot be shared. For this reason, we follow \cite{yildirim2014using} in using data for the US. The measurement of skill relatedness is based on O*Net data \cite{onet2013about}. O*Net records the use of hundreds of skills and tasks in hundreds of occupations \cite{onet2013production}. The labor flows between occupations in the US are derived from the Census Bureau's Current Population Survey\footnote{\url{http://www.census.gov/programs-surveys/cps.html}}. Both datasets are public and shared at the URL provided in Section \ref{sec:introduction}.

For each pair of occupations we have the number of workers who changed jobs from one occupation in 2009 to another occupation in 2010. Job switchers who did not change occupation are counted in the diagonal of the matrix. The skill relatedness between occupations is assessed as follows. O*Net provides estimates of the relevance of each skill and task to an occupation using both domain experts and worker surveys. The database links an occupation to a skill or task with two scores: how important the skill or task is to perform the occupation and at which level the skill must be mastered (expert, medium, beginner). We keep the occupation-skill association if both scores are higher than the average importance and level of the skill or task across all occupations. Next, the similarity of two occupations is derived from the number of skills the occupations have in common. This yields a weighted undirected co-occurrence network.

\begin{figure}
\centering
\includegraphics[width=.9\columnwidth]{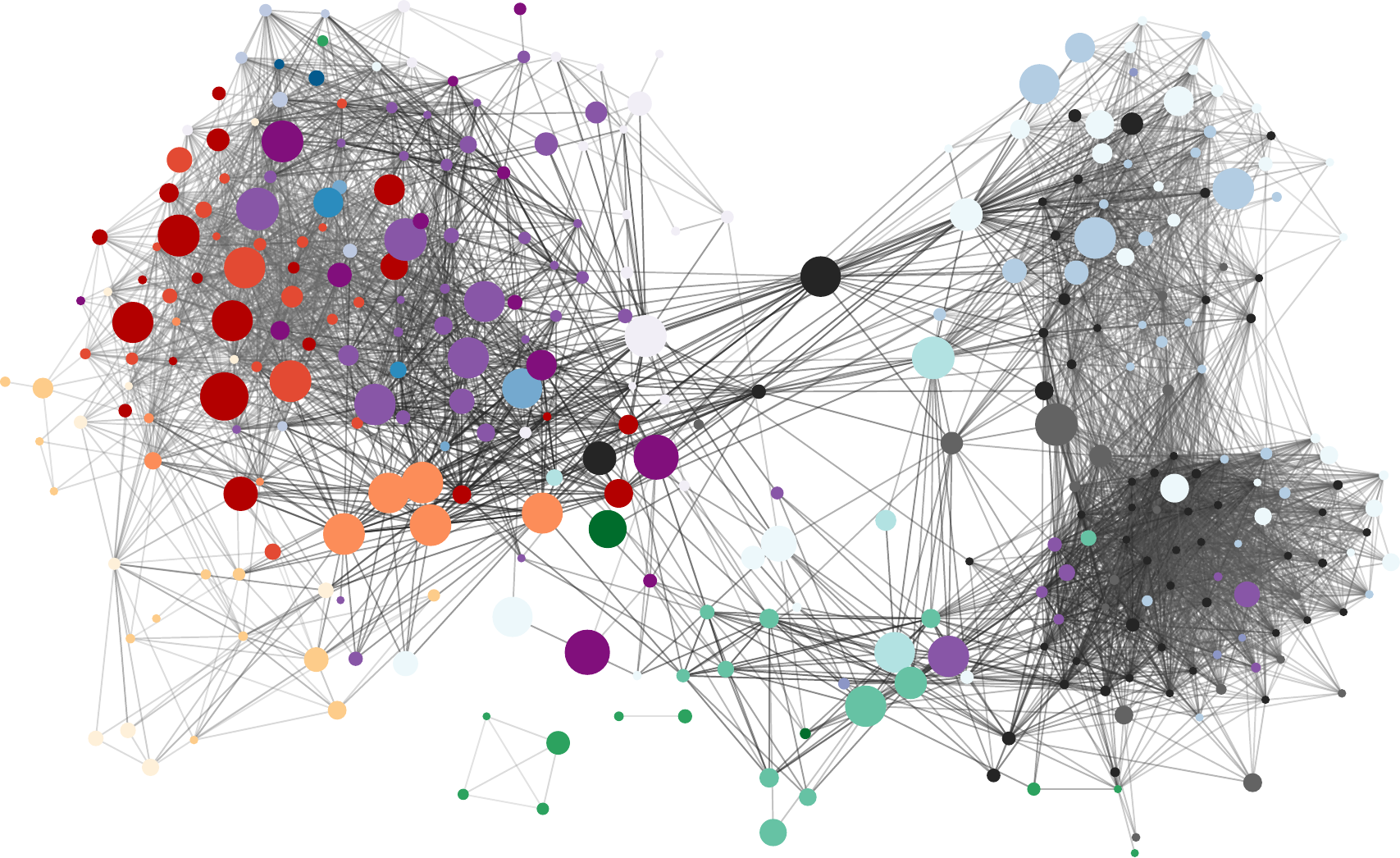}
\caption{The occupation co-occurence NC backbone.}
\label{fig:onet-neffke}
\end{figure}

\begin{figure}
\centering
\includegraphics[width=.72\columnwidth]{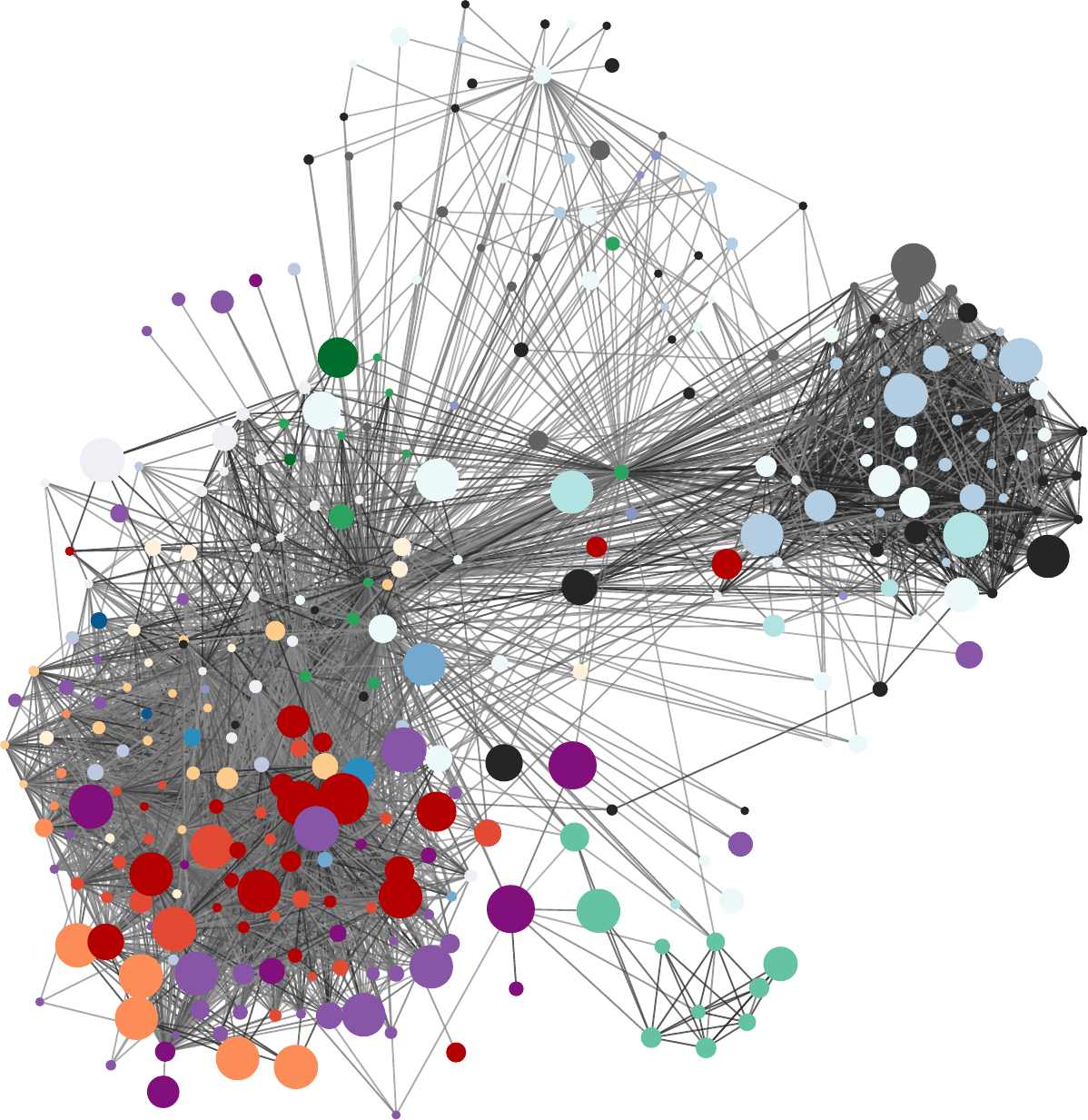}
\caption{The occupation co-occurence DF backbone.}
\label{fig:onet-vespignani}
\end{figure}

We calculate two backbones of this co-occurrence network. Figure \ref{fig:onet-neffke} depicts the backbone extracted with our NC method, while Figure \ref{fig:onet-vespignani} shows the one generated with the Disparity Filter. High Salience Skeleton and Doubly Stochastic are not reported, because the DS transformation was not possible, and the HSS could not generate the backbone in a reasonable time. Each node is an occupation. Nodes are connected if they have a significant number of skills and tasks in common, using the different backboning criteria to filter connections. The two networks have roughly the same number of connections. The nodes are colored using the first digit of the occupation classification. Node size is proportional to the amount of people working in an occupation. Edge colors are proportional to the significance of the link according to the backbone technique, from black (very significant) to gray (least significant).

Note how the DF backbone in Figure \ref{fig:onet-vespignani} looks, and indeed is, more dense. The reason is that the DF dropped more nodes than NC, around fifty -- a significant part of the network. This is further proof that NC methodology extracts backbones of a higher quality topology.

From the figures we can see that the NC backbone is able to capture an underlying structure in the data. The DF backbone has two dishomogeneous clusters, where seemingly connections are established between almost any pair of occupations. In the NC backbone, we can clearly identify a number of potential modules, which seem to be correlated with the occupation classification. 

This is not an artifact of the visualization: the Infomap community discovery algorithm \cite{rosvall2008maps} was able to compress the NC backbone with a codelength 15.0\% smaller than without communities (from 7.97 bits to 6.78), against the 9.3\% gain obtained with the DF backbone (from 7.69 bits to 6.98). Moreover, given that the classification is made by experts grouping occupations, occupations belonging to the same class should be related to one another. In line with this, the modularity \cite{newman2006modularity} of the partition using the first two digits of the occupation code as community is higher for the NC backbone (0.192) than for the DF backbone (0.115). Also the normalized mutual information between the backbone communities and the two digit occupation classification is higher for the NC backbone (0.423) than for the DF backbone (0.401). This means that the communities found by Infomap on the NC backbone are a better predictor of the first two digits of the occupation code than the communities found in the DF backbone.

Coming to the prediction task, our model assumes that the more skills two occupations have in common, the more people will switch between them. This is implemented in the following simple model:

$$F_{ij} = \beta_1 C_{ij} + \beta_2 S_{i \cdot} + \beta_3 S_{\cdot j} + \epsilon_{ij},$$

where $F_{ij}$ is the number of people switching from $i$ to $j$, $C_{ij}$ is the number of skills and tasks $i$ and $j$ have in common, and $S$ indicates the size of $i$ as an origin ($S_{i \cdot}$), and of $j$ as destination ($S_{\cdot j}$) in terms of the total number of occupation switches. We consider the direction of the switch, so $F_{ij}$ records together people moving $i \rightarrow j$, and $F_{j,i}$ records together people moving $j \rightarrow i$.

When testing this relationship using all $(i,j)$ pairs, we obtain a correlation of 0.390. When we restrict to only the $(i,j)$ pairs included in the DF backbone, we obtain a higher correlation, equal to 0.431. This means that the flows between the occupational pairs that were left after filtering out noisy connections are easier to predict. Our NC backbone increases the correlation further to 0.454. This means that the NC backbone was able to capture even better which pairs are similar both in terms of their skill relatedness and in the eyes of occupation switchers.

\section{Conclusion}
In this paper we focus on the problem of detecting the backbone of a complex network. A backbone is a subset of a network that contains the largest possible subset of nodes and the smallest possible subset of edges, such that the substantive and topological characteristics of the network are preserved. We focus on generic network backboning as a common first step in a network analysis task, without considering specialized applications. We compare our method mainly to the Disparity Filter and the High Salience Skeleton, which are the current state of the art in generic network backboning. Our Noise-Corrected  backbone starts from a null model in which edge weights are drawn from a binomial distribution. We estimate an expectation and variance of edge weights, simultaneously considering the propensity of the origin node to emit and the destination node to receive edges. This improves over the null model underlying the Disparity Filter, which considers origin and desitination nodes separately instead of bilaterally. Experiments show that the Noise-Corrected backbone performs well on three critical evaluation criteria. NC backbones can handle low and high amounts of noise, as shown in synthetic network experiments. NC backbones have comparable coverage and stability with the state of the art. NC backbones are of high quality, being able to improve the performance of edge weight predictive models, also in real-world scenarios, as our case study shows. Moreover, the methodology scales almost linearly in number of edges, scaling to million of edges in minutes.

This paper paves the way for further developments in network backboning. First, since we are introducing a new way to remove noise from network connections, we can investigate how noise affects classical network science results in depth. For instance, we plan to study whether it is possible to distinguish real from spurious changes in networks. If so, we can explore how noise impacts studies that look at contagion, shortest paths, clustering, and so on. Second, our method is at present only defined for networks with a single relationship type. We can extend the NC methodology to consider multilayer networks, where nodes in different layers are coupled together and where these couplings influence the backbone structure. Finally, we can explore improvements in the implementation of the NC backbone, exploiting optimizations that could lead to its potential application to networks with billions of edges.

\section*{Acknowledgment}
We thank Sebastian Bustos for cleaning the trade data; and Renaud Lambiotte, Michael Schaub, and Andres Gomez for helpful discussions. We thank the Mastercard Center for Inclusive Growth for providing access to their anonymized and aggregated transaction data. Michele Coscia has been partly supported by FNRS, grant \#24927961.

\bibliographystyle{IEEEtranS}

\bibliography{biblio}

\end{document}